\documentclass{article}%
\usepackage{amssymb}
\usepackage{amsfonts}
\usepackage{amsmath}
\usepackage{graphicx}
\usepackage{geometry}

\begin{document}

\begin{center}
{\LARGE Bi-Hamiltonian Structures of Chaotic Dynamical Systems in 3D}

\bigskip

\bigskip

\bigskip

O\u{g}ul Esen\footnote{oesen@yeditepe.edu.tr}$^{,a}$, Anindya Ghose
Choudhury\footnote{aghosechoudhury@gmail.com}$^{,b}$, Partha
Guha\footnote{partha@bose.res.in}$^{,c}$

\bigskip
\end{center}

$^{a}$Department of Mathematics, Yeditepe University, 34755 Ata\c{s}ehir,
Istanbul, Turkey,

$^{b}$Department of Physics, Surendranath College, 24/2 Mahatma Gandhi Road,
Calcutta 700009, India,

$^{c}$SN Bose National Centre for Basic Sciences, JD Block, Sector III, Salt
Lake, Kolkata 700098, India,

\bigskip

\textbf{Abstract:} We study Poisson structures of dynamical systems with three
degrees of freedom which are known for their chaotic properties, namely
L\"{u}, modified L\"{u}, Chen, $T$ and Qi systems. We show that all these
flows admit bi-Hamiltonian structures depending on the values of their parameters.

\textbf{MSC(2010): }34C14, 34C20.

\textbf{Key words: }First integrals, Poisson structures, Jacobi's last
multiplier, chaotic three dimensional systems, bi-Hamiltonian structures,
Nambu-Poisson bracket.

\bigskip

\section{Introduction}

In 1963, Lorenz discovered the first (autonomous and three dimensional)
chaotic dynamical system \cite{Lorenz}. Fifteen years later, R\"{o}ssler
constructed a more simple chaotic system involving only one nonlinear term
\cite{Ross}. More recently, in 1999, Chen and Ueta defined three dimensional
chaotic Chen system as a dual of Lorenz system using chaotification approach
\cite{CU1, CU2}. Three years after the presentation of the Chen system, L\"{u}
system was introduced \cite{LuChen,LuChen1}. L\"{u} system is an intermediate
system that bridges the gap between the Chen and the Lorenz systems. In 2005,
chaotic Qi system was presented by completing quadratic cross product terms
some of which are missing in the Chen system \cite{Qi}. In 2008, Tigan
presented chaotic $T$ system which is structurally looks like the same with
the Lorenz system except a linear term \cite{T1}.

Even though the association of chaotic behaviour with integrability had not
generally been expected, many extensive studies have been done on the
Hamiltonian structures of chaotic dynamics. Two of these studies,
\cite{GuNu93} and \cite{Gao}, are important for the present paper. In
\cite{GuNu93}, G\"{u}mral and Nutku showed that every Hamiltonian system in
three dimensions is mutually bi-Hamiltonian and derived non-canonical
bi-Hamiltonian structure of the Lorenz system for certain values of its
parameters. In \cite{Gao}, Gao showed that the necessary and sufficient
condition of a three dimensional dynamical system $\dot{\mathbf{x}}%
=\mathbf{X}$ (having a time independent first integral) being Hamiltonian is
existence of a non-zero function $M$, called Jacobi's last multiplier, which
makes $M\mathbf{X}$ divergence free. Using this, Gao achieved to present
Hamiltonian structure of the Lotka-Volterra system. In the present work,
motivating by these observations and following historical order presented in
the previous paragraph, we shall investigate the Hamiltonian structures of
three dimensional chaotic L\"{u}, modified L\"{u}, Chen, Qi and $T$ systems.
We shall able to show that all these systems are bi-Hamiltonian for certain
values of their parameters.

This paper consists of two main sections. In order to make the paper
self-contained, next section will be started with the definitions of Poisson
and Nambu-Poisson brackets, and presentations of Hamiltonian, bi-Hamiltonian
and Nambu-Hamiltonian systems. We shall point out three important properties
of the Hamiltonian structures in dimension three. The first one is that they
always come in compatible pairs to form a bi-Hamiltonian structure
\cite{GuNu93}. The second one is that existence of a Jacobi's last multiplier
$M$ completely determines existence of bi-Hamiltonian structure of a dynamical
system \cite{Gao}. The third one is that, in three dimensions, a Hamiltonian
(hence a bi-Hamiltonian) system is mutually Nambu-Hamiltonian.

In section $3$, we shall focus on three-dimensional autonomous L\"{u},
modified L\"{u}, Chen, Qi and $T$ systems. One common feature of these chaotic
systems is that, for some certain values of their parameters, they can be
transformed in such a way that the generating vector fields become divergence
free. Hence, Jacobi's last multipliers of all these systems can be taken as
the unity. This enables us to derive bi-Hamiltonian structures of them after
the presentation of a\ time independent first integral. We remark that, for
all these systems, except the L\"{u} system, one of the Poisson structures is time-dependent.

This work may be considered as a complementary to \cite{EsGhGuGu15}, where
bi-Hamiltonian structures and superintegrability of some four dimensional
hyperchaotic systems were studied.

\section{Bi-Hamiltonian structure of 3D systems}

\subsection{Poisson brackets}

A first integral of a dynamical system is a real-valued function that retains
constant values on integral curves of the system. An $n$-dimensional system is
maximally superintegrable if it admits $n-1$ first integrals
\cite{Gor01,TeWiHaMiPoRo}. Note that, existence of $n-1$ first integrals lets
one to reduce the systems of differential equations to a one quadrature. In
dimension three, two first integrals are needed for (maximal) superintegrability.

Poisson bracket on an $n$-dimensional space is a binary operation
$\{\bullet,\bullet\}$ on the space of real-valued smooth functions satisfying
the Leibnitz and the Jacobi identities \cite{LaPi12,LM,OLV,wei83}. We define a
Poisson bracket of two functions $F$ and $H$ by
\begin{equation}
\left\{  F,H\right\}  =\nabla F\cdot N\nabla H, \label{PB}%
\end{equation}
where $N$ is skew-symmetric Poisson matrix, $\nabla F$ and $\nabla H$ are
gradients of $F$ and $H$, respectively. Poisson bracket (\ref{PB})
automatically satisfies the Leibnitz identity, whereas, in a local frame
$\left(  x^{i}\right)  $, the Jacobi identity turns out to be%
\begin{equation}
N^{i[j}\partial_{x^{i}}N^{kl]}=0 \label{jcb}%
\end{equation}
assuming summation over repeated indices. In Eq.(\ref{jcb}), $N^{ij}$ are
components of the Poisson matrix $N$, superscript bracket $[$ $\ ]$ refers
anti-symmetrization, and $\partial_{x^{i}}$ denotes the partial
differentiation with respect to $x^{i}$. A Casimir function $C$ on a Poisson
space is the one that commutes with all the other functions. In order to have
a non-trivial Casimir function, the Poisson matrix $N$ must be degenerate.

A system of ODE's is Hamiltonian if it can be written in the form of
Hamilton's equation%
\begin{equation}
\dot{x}^{i}=\left\{  x^{i},H\right\}  \text{, \ \ }i=1,...,n, \label{HamEqn}%
\end{equation}
for $H$ being a real-valued function, called Hamiltonian function, and
$\{\bullet,\bullet\}$ being a Poisson bracket. For the Poisson bracket
(\ref{PB}), Hamilton's equation takes the particular form
\begin{equation}
\mathbf{\dot{x}}=N\nabla H. \label{HamEq}%
\end{equation}
In order to express a finite-dimensional dynamical system in the form of
Hamilton's equation, two tasks must be accomplished. One is to define a proper
Poisson matrix and the other is to find a proper Hamiltonian function. When
searching a Poisson matrix, main hurdle is to find a solution of the Jacobi
identity (\ref{jcb}). For three dimensional flows, this can be computed
relatively simply since the Jacobi identity yields a scalar equation. On the
other hand, in order to find a Hamiltonian function, one may try to find the
first integrals of the system, because it is often observed that a first
integral or some function of the first integral(s) becomes the Hamiltonian
function. Note that, for autonomous systems, in case of the existences, both
of the Hamiltonian and the Casimir functions are the first integrals of the
system \cite{Gor01,OLV}, whereas, for the case of non-autonomous systems, they
may depend explicitly on time so that they may fail to be integral invariants
\cite{AbMa78}.

A dynamical system is bi-Hamiltonian if it admits two different Hamiltonian
structures
\begin{equation}
\mathbf{\dot{x}}=N_{1}\nabla H_{2}=N_{2}\nabla H_{1}, \label{biHam}%
\end{equation}
with the requirement that the Poisson matrices $N_{1}$ and $N_{2}$ be
compatible, that is any linear pencil $N_{1}+cN_{2}$ must satisfy the Jacobi
identity (\ref{jcb}), \cite{MaMo84,OLV}. In three dimensions, an autonomous
bi-Hamiltonian system is superintegrable. We refer \cite{GN} for the
multi-Hamiltonian structures of the maximal superintegrable systems of
arbitrary order.

\subsection{Nambu-Poisson brackets}

In \cite{Nambu}, a ternary operation $\{\bullet,\bullet,\bullet\}$, called
Nambu-Poisson bracket, is defined on the space of smooth functions satisfying
the generalized Leibnitz identity
\begin{equation}
\left\{  F_{1},F_{2},FH\right\}  =\left\{  F_{1},F_{2},F\right\}  H+F\left\{
F_{1},F_{2},,H\right\}  \label{GLI}%
\end{equation}
and the fundamental (or Takhtajan) identity
\begin{equation}
\left\{  F_{1},F_{2},\{H_{1},H_{2},H_{3}\}\right\}  =\sum_{k=1}^{3}%
\{H_{1},...,H_{k-1},\{F_{1},F_{2},H_{k}\},H_{k+1},...,H_{3}\}, \label{FI}%
\end{equation}
for arbitrary functions $F,F_{1},F_{2},H,H_{1},H_{2}$. A dynamical system is
called Nambu-Hamiltonian with Hamiltonian functions $H_{1}$ and $H_{2}$ if it
can be recasted as%
\begin{equation}
\dot{x}^{i}=\left\{  x^{i},H_{1},H_{2}\right\}  . \label{NHamEqn}%
\end{equation}

By fixing the Hamiltonian functions $H_{1}$ and $H_{2}$, we can write
Nambu-Hamiltonian system (\ref{NHamEqn}) in the bi-Hamiltonian form
\begin{equation}
\dot{x}^{i}=\left\{  x^{i},H_{1}\right\}  ^{H_{2}}=\left\{  x^{i}%
,H_{2}\right\}  ^{H_{1}}\label{NH-2Ham}%
\end{equation}
where the Poisson brackets $\{\bullet,\bullet\}^{H_{2}}$ and $\{\bullet
,\bullet\}^{H_{1}}$ are defined by
\begin{equation}
\left\{  F,H\right\}  ^{{H}_{2}}=\left\{  F,H,H_{2}\right\}  \text{,
\ \ \ }\left\{  F,H\right\}  ^{{H}_{1}}=\left\{  F,H_{1},H\right\}
,\label{Pois}%
\end{equation}
respectively. Both of the brackets $\{\bullet,\bullet\}^{H_{2}}$ and
$\{\bullet,\bullet\}^{H_{1}}$ satisfy the Jacobi and Leibnitz identities
\cite{Guha06}. In order to show that they constitute a compatible Poisson pair
we need to confirm that an arbitrary linear pencil $\{\bullet,\bullet
\}_{LP}=\{\bullet,\bullet\}^{H_{2}}+\lambda\{\bullet,\bullet\}^{H_{1}}$ is
satisfying the Jacobi identity. To establish this, we compute
\begin{align}
&  \underset{H,F,K}{\circlearrowleft}\left\{  \{H,F\}_{LP},K\right\}
_{LP}\nonumber\\
&  =\underset{H,F,K}{\circlearrowleft}(\left\{  \{H,F\}^{H_{2}}+\lambda
\{H,F\}^{H_{1}},K\right\}  ^{H_{2}}+\lambda\left\{  \{H,F\}^{H_{2}}%
+\lambda\{H,F\}^{H_{1}},K\right\}  ^{H_{1}})\nonumber\\
&  =\underset{H,F,K}{\circlearrowleft}(\{\left\{  H,F\}^{H_{2}},K\right\}
^{H_{2}}+\lambda\left\{  \{H,F\}^{H_{1}},K\right\}  ^{H_{2}}+\lambda\left\{
\{H,F\}^{H_{2}},K\right\}  ^{H_{1}}\nonumber\\
&  \text{ \ \ \ \ \ \ \ \ \ \ \ }+\lambda^{2}\left\{  \{H,F\}^{H_{1}%
},K\right\}  ^{H_{1}})\nonumber\\
&  =\lambda\underset{H,F,K}{\circlearrowleft}(\left\{  \{H,F\}^{H_{1}%
},K\right\}  ^{H_{2}}+\left\{  \{H,F\}^{H_{2}},K\right\}  ^{H_{1}})\nonumber\\
&  =\lambda\underset{H,F,K}{\circlearrowleft}(\left\{  \{H,H_{1}%
,F\},K,H_{2}\right\}  +\left\{  \{H,F,H_{2}\},H_{1},K\right\}  ),\label{la}%
\end{align}
where $\circlearrowleft$ denotes the sum over the cyclic permutation of $H,F$
and $K$ while keeping $H_{1}$ and $H_{2}$ constants. In the calculation, the
first and the fourth terms in second line vanish since the brackets
$\{\bullet,\bullet\}^{H_{2}}$ and $\{\bullet,\bullet\}^{H_{1}}$ satisfy the
Jacobi identity. In the fourth line, the definitions in Eq.(\ref{Pois}) are
used. In Eq.(\ref{la}), expansion of the summation, reordering the arguments
and application of the Takhtajan's identity (\ref{FI}) result with
\begin{align}
&  \lambda\left\{  \{H,H_{1},F\},K,H_{2}\right\}  +\lambda\left\{
\{F,H_{1},K\},H,H_{2}\right\}  +\lambda\left\{  \{K,H_{1},H\},F,H_{2}\right\}
\nonumber\\
&  \text{ \ \ \ \ \ \ \ \ \ \ \ }+\left\{  \{H,F,H_{2}\},H_{1},K\right\}
+\left\{  \{F,K,H_{2}\},H_{1},H\right\}  +\left\{  \{K,H,H_{2}\},H_{1}%
,F\right\}  \nonumber\\
&  =-\lambda\left\{  \{F,H_{1},H\},K,H_{2}\right\}  -\lambda\left\{
H,\{F,H_{1},K\},H_{2}\right\}  -\lambda\left\{  \{K,H,H_{1}\},F,H_{2}\right\}
\nonumber\\
&  \text{ \ \ \ \ \ \ \ \ \ \ \ }-\lambda\left\{  H_{1},\{F,H_{2}%
,H\},K\right\}  -\lambda\left\{  H_{1},H,\{F,H_{2},K\}\right\}  +\lambda
\left\{  \{K,H,H_{2}\},F,H_{1}\right\}  \nonumber\\
&  =-\lambda\left\{  F,H_{1},\{H,K,H_{2}\}\right\}  +\lambda\left\{
H,K,\{F,H_{1},H_{2}\}\right\}  -\lambda\left\{  \{K,H,H_{1}\},F,H_{2}\right\}
\nonumber\\
&  \text{ \ \ \ \ \ \ \ \ \ \ \ }-\lambda\left\{  F,H_{2},\{H_{1}%
,H,K\}\right\}  +\lambda\left\{  \{F,H_{2},H_{1}\},H,K\right\}  +\lambda
\left\{  \{K,H,H_{2}\},F,H_{1}\right\}  \nonumber\\
&  =0.
\end{align}
That is, $\{\bullet,\bullet\}_{LP}$ is a Poisson bracket, hence $\{\bullet
,\bullet\}^{H_{2}}$ and $\{\bullet,\bullet\}^{H_{1}}$ are compatible.
Inversely, in dimension three, expressing a bi-Hamiltonian dynamics in the
form of a Nambu-Hamiltonian system is possible as we shall show in the next
section. For a general discussion on the linear pencils of Poisson brackets,
we refer \cite{CaMaPe93} which is, additionally, showing how Casimir functions
of the pencil yield pairwise commuting functions with respect to both of the
brackets defining the pencil.

A generalization of Nambu-Poisson bracket was presented in \cite{Ta} by
introducing an $r$-ary operation, called generalized Nambu bracket, for
$r=2,3,...$. This generalization covers both of the Poisson and the
Nambu-Poisson brackets, that is, it reduces to a Poisson bracket when $r=2$,
and it turns out to be a classical Nambu bracket when $r=3$. A system is
generalized Nambu-Hamiltonian if it can be written as
\begin{equation}
\dot{x}^{i}=\left\{  x^{i},H_{1},H_{2},...,H_{r-1}\right\}
\end{equation}
under the existence of $r-1$ number of Hamiltonian functions. Note that, by
generalizing the procedure described in Eq.(\ref{NH-2Ham}), a generalized
Nambu-Hamiltonian system can be recasted as a multi-Hamiltonian system. For a
detailed discussion on the relationship between generalized Nambu-Poisson and
Poisson brackets, we refer \cite{Guha06, TeVe04}. In \cite{Guha06},
additionally, investigations on quadratic Poisson structures containing
celebrated Sklyanin algebras can be found.

\subsection{Bi-Hamiltonian structures in three dimensions}

Space of three dimensional vectors and space of three by three skew-symmetric
matrices are isomorphic via the map
\begin{equation}
\mathbf{J}=\left(  X,Y,Z\right)  \longleftrightarrow N={\left(
\begin{array}
[c]{ccc}%
0 & -Z & Y\\
Z & 0 & -X\\
-Y & X & 0
\end{array}
\right)  .} \label{iso}%
\end{equation}
The isomorphism (\ref{iso}) can be realized by defining the identity
$N\mathbf{B}=\mathbf{J}\times\mathbf{B}$ for $\mathbf{B}$ being an arbitrary
vector. Existence of this isomorphism enables us to identify a three by three
Poisson matrix $N$ with a three dimensional Poisson vector field $\mathbf{J}$.
In this case, the Jacobi identity (\ref{jcb}) turns out to be%
\begin{equation}
\mathbf{J}\cdot(\nabla\times\mathbf{J})=0, \label{jcbv}%
\end{equation}
see, for example, \cite{Gum1}. $C$ is a Casimir function if its gradient
$\nabla C$ is parallel to $\mathbf{J}$ that is
\begin{equation}
\mathbf{J}\times\nabla C=\mathbf{0}. \label{CasimirC}%
\end{equation}

Under the isomorphism (\ref{iso}), in three dimensions, Hamilton's equation
takes the particular form%
\begin{equation}
\mathbf{\dot{x}}=\mathbf{J}\times\nabla H, \label{HamEq3}%
\end{equation}
whereas a bi-Hamiltonian system is in form
\begin{equation}
\mathbf{\dot{x}}=\mathbf{J}_{1}\times\nabla H_{2}=\mathbf{J}_{2}\times\nabla
H_{1}. \label{bi-Ham}%
\end{equation}
Note that, in order to guarantee the compatibility, Poisson vector fields
$\mathbf{J}_{1}$ and $\mathbf{J}_{2}$ must satisfy
\begin{equation}
\mathbf{J}_{1}\cdot\nabla\times\mathbf{J}_{2}+\mathbf{J}_{2}\cdot\nabla
\times\mathbf{J}_{1}=0. \label{J}%
\end{equation}

General solution of the Jacobi identity (\ref{jcbv}) is
\begin{equation}
\mathbf{J}=\frac{1}{M}\nabla H_{1} \label{Nsoln}%
\end{equation}
for arbitrary functions $M$ and $H_{1}$ \cite{AGZ,HB1,HB2,HB3}. Existence of
the scalar multiple $1/M$ in the solution is a manifestation of the conformal
invariance of Jacobi identity. In the literature, $M$ is called Jacobi's last
multiplier \cite{Gor01,Jac1,Jac2,Whi}. The potential function $H_{1}$ in
Eq.(\ref{Nsoln}) is a Casimir function of the Poisson vector field
$\mathbf{J}$. Any other Casimir of $\mathbf{J}$ has to be linearly dependent
to the potential function$\ H_{1}$ since the kernel of the equation
(\ref{CasimirC}) is one dimensional. Substitution of the general solution
(\ref{Nsoln}) of $\mathbf{J}$ into the Hamilton's equation (\ref{HamEq3})
results with
\begin{equation}
\dot{\mathbf{x}}=\frac{1}{M}\nabla H_{1}\times\nabla H_{2}. \label{x1}%
\end{equation}
This shows that velocity vector $\dot{\mathbf{x}}$ is orthogonal to the
gradient functions, that is the solution curve $\mathbf{x}$ is parallel to the
intersection of level surfaces of $H_{1}$ and $H_{2}$ \cite{GC}. In the rest
of this subsection, we investigate geometry of the Hamilton's equation
(\ref{x1}) in detail.

The first observation is as follows. We choose two Poisson vector fields
\begin{equation}
\mathbf{J}_{1}=\frac{1}{M}\nabla H_{1}\text{, \ \ }\mathbf{J}_{2}=-\frac{1}%
{M}\nabla H_{2}. \label{J12}%
\end{equation}
Substitution of the Poison vector fields in Eq.(\ref{J12}) into the
compatibility condition (\ref{J}) results with%
\begin{align}
-\frac{1}{M}  &  \nabla H_{1}\cdot\nabla\times(\frac{1}{M}\nabla H_{2}%
)-\frac{1}{M}\nabla H_{2}\cdot\nabla\times\frac{1}{M}\nabla H_{1}\nonumber\\
&  =-\frac{1}{M^{2}}\nabla H_{1}\cdot(\nabla\times\nabla H_{2})+\frac{1}%
{M^{3}}\nabla H_{1}\cdot(\nabla M\times\nabla H_{2})-\frac{1}{M^{2}}\nabla
H_{2}\cdot(\nabla\times\nabla H_{1})\nonumber\\
&  \text{ \ \ \ \ \ \ \ \ \ \ \ \ }+\frac{1}{M^{3}}\nabla H_{2}\cdot(\nabla
M\times\nabla H_{1})\nonumber\\
&  =0,
\end{align}
where the skew-symmetry of the triple product is used. So that, Poisson vector
fields in Eq.(\ref{J12}) are compatible and Hamilton's equation (\ref{x1}) is
bi-Hamiltonian in the form of Eq.(\ref{bi-Ham}). This shows that Hamiltonian
structures of dynamical systems, in three dimensions, always come in
compatible pairs to form a bi-Hamiltonian structure \cite{GuNu93}.

The second observation can be stated as follows. A three dimensional dynamical
system $\dot{\mathbf{x}}=\mathbf{X}$ having a time independent first integral
is bi-Hamiltonian if and only if there exist a Jacobi's last multiplier $M$
which makes $M\mathbf{X}$ divergence free \cite{Gao}. Sketch of the proof is
as follows. For a bi-Hamiltonian dynamical system $\dot{\mathbf{x}}%
=\mathbf{X}$ in the form of Eq.(\ref{x1}), scalar multiple $M\mathbf{X}$ of
the vector field $\mathbf{X}$ equals to the cross product $\nabla H_{1}%
\times\nabla H_{2}$ of two gradients. Hence, the divergence of $M\mathbf{X}$
vanishes, that is
\begin{equation}
\nabla\cdot\left(  M\mathbf{X}\right)  =0. \label{JLM}%
\end{equation}
Inversely, assume that a dynamical system $\dot{\mathbf{x}}=\mathbf{X}$ (not
necessarily Hamiltonian) has a first integral $C$. Then the gradient $\nabla
C$ is parallel to the vector field $\mathbf{X}$ hence parallel to any of its
scalar multiple, say $M\mathbf{X}$. There are infinitely many vectors
satisfying
\begin{equation}
M\mathbf{X}=\nabla C\times\mathbf{A}. \label{B}%
\end{equation}
We choose the vector field $\mathbf{A}$ so that angle between the curl
$\nabla\times\mathbf{A}$ and $\nabla C$ be different than $\pi/2$. The
divergence condition (\ref{JLM}) becomes%
\begin{align}
0  &  =\nabla\cdot\left(  M\mathbf{X}\right)  =\nabla\cdot\left(  \nabla
C\times\mathbf{A}\right) \nonumber\label{A}\\
&  =\nabla\times\nabla C+\nabla C\cdot\nabla\times\mathbf{A}\nonumber\\
&  =\nabla C\cdot\nabla\times\mathbf{A}.
\end{align}
This is valid only if the curl of $\mathbf{A}$ vanishes. Thus, $\mathbf{A}$
locally equals to a gradient vector field $\nabla H$ for some function $H$.
Substitution of $\nabla H$ in Eq.(\ref{B}) enables us to recast $\dot
{\mathbf{x}}=\mathbf{X}$ in the form of Eq.(\ref{x1}). For the first
Hamiltonian formulation, we take $H$ as the Hamiltonian function and $\left(
1/M\right)  \nabla C$ as the Poisson vector field. For the second Hamiltonian
formulation, we take $C$ as the Hamiltonian function and $-\left(  1/M\right)
\nabla H$ as the Poisson vector.

When, particularly, a system $\dot{\mathbf{x}}=\mathbf{X}$ is divergence-free,
then any constant function can be taken as the Jacobi's last multiplier. In
this case, finding a time independent first integral is sufficient to conclude
that the system is bi-Hamiltonian. For the use of method of Jacobi's last
multiplier to determine Lagrangian picture of dynamical systems, we refer
\cite{CGK,NL1,NL2}.

The third observations is as follows. On the three dimensional space, we
define a bracket of three functions $F$, $H_{1}$ and $H_{2}$ as the triple
product
\begin{equation}
\left\{  F,H_{1},H_{2}\right\}  =\frac{1}{M}\nabla F\cdot\nabla H_{1}%
\times\nabla H_{2} \label{NambuPois}%
\end{equation}
of their gradient vectors. It is immediate to see that the bracket in
Eq.(\ref{NambuPois}) satisfies both of the generalized Leibnitz identity
(\ref{GLI}) and the fundamental identity (\ref{FI}), so that it is a
Nambu-Poisson bracket. Note that, the Hamilton's equation (\ref{x1}) is
Nambu-Hamiltonian (\ref{NHamEqn}) with the bracket (\ref{NambuPois}) having
the Hamiltonian functions $H_{1}$ and $H_{2}$. Combining this observation with
the first one, we conclude that, a Hamiltonian (hence a bi-Hamiltonian)
dynamics in three dimensions is mutually Nambu-Hamiltonian whereas combining
with the second one, we conclude that, a system is Nambu-Hamiltonian if and
only if there exists a Jacobi's last multiplier satisfying the divergence
condition in Eq.(\ref{JLM}).

\section{Examples}

\subsection{L\"{u} system}

Chaotic L\"{u} system consists of three autonomous first order differential
equations
\begin{align}
\dot{x}  &  =\alpha(y-x),\label{Lu}\\
\dot{y}  &  =\gamma y-xz,\nonumber\\
\dot{z}  &  =xy-\beta z,\nonumber%
\end{align}
where $\alpha,\beta$ and $\gamma$ are real constant parameters
\cite{LuChen,LuChen1}. We, particularly, take $\beta=-\gamma=2\alpha$ and
change the dependent variables $\left(  x,y,z\right)  $ to $\left(
u,v,w\right)  $ according to
\begin{equation}
u=xe^{\alpha t},\;\ \;v=ye^{-\gamma t},\;\;\;w=ze^{\beta t}.
\end{equation}
Additionally, we rescale the time variable by $\bar{t}=-e^{-\alpha t}/\alpha$.
Then L\"{u} system (\ref{Lu}) turns out to be
\begin{equation}
u^{\prime}=\alpha v,\;\;\;v^{\prime}=-uw,\;\;\;w^{\prime}=-uv, \label{Lu4}%
\end{equation}
where, prime denotes derivative with respect to the new time parameter
$\bar{t}$.

The system (\ref{Lu4}) is divergence free, hence for the satisfaction of the
PDE in (\ref{JLM}), Jacobi's last multiplier $M$ must a constant function. In
this case, the system (\ref{Lu4}) admits two time independent first integrals
\begin{equation}
H_{1}(u,v,w)=\frac{1}{2}(v^{2}+w^{2})\text{, \ \ }H_{2}(u,v,w)=\frac{1}%
{2}u^{2}-\alpha w. \label{Lu5}%
\end{equation}
The definitions of Poisson vectors in Eq.(\ref{J12}) enables us to arrive
\begin{equation}
\mathbf{J}_{1}=\nabla H_{1}=\left(  0,v,w\right)  \text{, \ \ }\mathbf{J}%
_{2}=-\nabla H_{2}=\left(  -u,0,\alpha\right)  . \label{LuPois}%
\end{equation}
So that, the system (\ref{Lu4}) exhibits a bi-Hamiltonian structure in form of
Eq.(\ref{bi-Ham}) by choosing Hamiltonian functions $H_{1}$ and $H_{2}$ as in
Eq.(\ref{Lu5}) and choosing Poisson vectors $\mathbf{J}_{1}$ and
$\mathbf{J}_{2}$ as in Eq.(\ref{LuPois}), respectively.

\subsection{A modified L\"{u} system}

We add a cross product term $yz$ to the right hand side of the first equation
in the L\"{u} system (\ref{Lu}) and obtain
\begin{align}
\dot{x}  &  =\alpha(y-x)+yz,\label{LuM1a}\\
\dot{y}  &  =\gamma y-xz,\nonumber\\
\dot{z}  &  =xy-\beta z,\nonumber
\end{align}
which is called modified L\"{u} system, see \cite{Radio}. We take
$\beta=2\alpha$ and $\gamma=\alpha$, then apply a change of variables
\begin{equation}
u=xe^{\alpha t},\qquad v=ye^{-\alpha t},\qquad w=ze^{2\alpha t},
\end{equation}
which results with a non-autonomous system
\begin{equation}
\dot{u}=\alpha v+vwe^{-2\alpha t},\qquad\dot{v}=-uwe^{-2\alpha t},\qquad
\dot{w}=uv, \label{LuM}%
\end{equation}
admitting a time independent first integral
\begin{equation}
H_{1}(u,v,w)=\frac{1}{2}(u^{2}+v^{2})-\alpha w. \label{ModLuInt1}%
\end{equation}
The system \eqref{LuM} is divergence free hence we take Jacobi' last
multiplier as $M=1$ and arrive the following Poisson vector field
\begin{equation}
\mathbf{J}_{1}=\nabla H_{1}=\left(  u,v,-\alpha\right)  . \label{ModLuJ1}%
\end{equation}
with Hamiltonian function
\begin{equation}
H_{2}=-\frac{v^{2}}{2}+\frac{e^{2\alpha t}}{2\alpha^{2}}((u^{2}+v^{2}%
)(\frac{1}{4}(u^{2}+v^{2})-\alpha w)). \label{ModLuInt2}%
\end{equation}
Note that, $H_{2}$ is time dependent hence it is not an integral invariant of
the system (\ref{LuM}). Using $H_{2}$, we construct a time dependent Poisson
matrix
\begin{equation}
\mathbf{J}_{2}=-\nabla H_{2}=\left(  -\frac{e^{2\alpha t}u}{\alpha^{2}}\left(
u^{2}+v^{2}-\alpha v\right)  ,v-\frac{e^{2\alpha t}v}{\alpha^{2}}\left(
v^{2}+u^{2}-\alpha u\right)  ,\frac{e^{2\alpha t}}{2\alpha}(u^{2}%
+v^{2})\right)  . \label{ModLuJ2}%
\end{equation}
As a result, we proved that the modified L\"{u} system (\ref{LuM}) is
bi-Hamiltonian in form of Eq.(\ref{bi-Ham}) with Hamiltonian functions $H_{1}$
and $H_{2}$ in Eqs.(\ref{ModLuInt1}) and (\ref{ModLuInt2}), and Poisson
vectors $\mathbf{J}_{1}$ and $\mathbf{J}_{2}$ in Eqs.(\ref{ModLuJ1}) and
(\ref{ModLuJ2}), respectively.

\subsection{The $T$-system}

Explicitly, $T$-system is given by
\begin{align}
\dot{x}  &  =\alpha(y-x),\label{tsys1}\\
\dot{y}  &  =(\gamma-\alpha)x-\alpha xz,\nonumber\\
\dot{z}  &  =xy-\beta z,\nonumber
\end{align}
with $\alpha,\beta,\gamma$ being real parameters for $\alpha$ being non-zero
\cite{T1}. When $\beta=2\alpha$, the system (\ref{tsys1}) admits a
time-dependent first integral
\begin{equation}
H_{1}=e^{2\alpha t}(x^{2}-2\alpha z). \label{tsys2}%
\end{equation}

Under the change of variables
\begin{equation}
u=xe^{\alpha t},\;\;v=y,\;\;w=ze^{2\alpha t} \label{tsys3}%
\end{equation}
the $T$-system (\ref{tsys1}) is transformed to a non-autonomous system
\begin{equation}
\dot{u}=\alpha ve^{\alpha t},\;\;\;\dot{v}=(\gamma-\alpha)ue^{-\alpha
t}-\alpha uwe^{-3\alpha t},\;\;\;\dot{w}=uve^{\alpha t} \label{tsys4}%
\end{equation}
whereas the first integral $H_{1}$ in Eq.(\ref{tsys2}) turns out to be time
independent function%
\begin{equation}
H_{1}=u^{2}-2\alpha w. \label{THam1}%
\end{equation}
The Jacobi's last multiplier for the set of equations (\ref{tsys4}) is $M=1$.
Hence, we compute Poisson vector field
\begin{equation}
\mathbf{J}_{1}=\nabla H_{1}=\left(  2u,0,-2\alpha\right)  \label{TJ1}%
\end{equation}
and Hamiltonian function
\begin{equation}
H_{2}=\frac{1}{2\alpha}\left[  \frac{1}{8}e^{-3\alpha t}u^{4}+\frac{1}%
{2}(\gamma-\alpha)e^{-\alpha t}u^{2}-\frac{\alpha}{2}u^{2}we^{-3\alpha
t}\right]  -\frac{1}{4}v^{2}e^{\alpha t}. \label{THam2}%
\end{equation}
Note that, the Hamiltonian function $H_{2}$ in Eq.(\ref{THam2}) is
non-autonomous hence not an integral invariant of the motion. Using $H_{2}$,
we construct a second Poisson vector
\begin{equation}
\mathbf{J}_{2}=-\nabla H_{2}=\left(  -\frac{1}{4\alpha}u^{3}e^{-3\alpha
t}+\frac{1}{2\alpha}(\gamma-\alpha)ue^{-\alpha t}+\frac{1}{2}uwe^{-3\alpha
t},\frac{v}{2}e^{\alpha t},\frac{1}{4}u^{2}e^{-3\alpha t}\right)  .
\label{TJ2}%
\end{equation}
The $T$-system given in Eq.(\ref{tsys4}) can be recasted as a bi-Hamiltonian
form as in Eq.(\ref{bi-Ham}) where the Hamiltonian functions $H_{1}$ and
$H_{2}$ are in Eqs.(\ref{THam1}) and (\ref{THam2}), and Poisson vectors
$\mathbf{J}_{1}$ and $\mathbf{J}_{2}$ are in Eqs.(\ref{TJ1}) and (\ref{TJ2}), respectively.

\subsection{The Chen system}

Chen system
\begin{align}
\dot{x}  &  =\alpha(y-x),\label{Chensys1}\\
\dot{y}  &  =(\gamma-\alpha)x+\gamma y-xz,\nonumber\\
\dot{z}  &  =xy-\beta z,\nonumber
\end{align}
is a chaotic system obtained by adding a state feedback to the second equation
of the Lorenz system \cite{CU1,CU2}. The Chen system (\ref{Chensys1}) is
topologically inequivalent to the Lorenz system and may be considered as the
dual of the Lorenz system in the sense defined by Van$\breve{e}\breve{c}$ek
and C$\breve{e}$likovsk$\acute{y}$ \cite{CV}. When $\beta=2\alpha$ it admits a
first integral
\begin{equation}
H_{1}=e^{2\alpha t}(x^{2}-2\alpha z). \label{ChenInt1}%
\end{equation}

Employing the change of variables
\begin{equation}
u=xe^{\alpha t},\;\;v=ye^{-\gamma t},\;\;w=ze^{2\alpha t}, \label{Chensys3}%
\end{equation}
the system (\ref{Chensys1}) may be expressed as
\begin{equation}
\dot{u}=\alpha ve^{(c+\alpha)t},\;\;\;\dot{v}=(\gamma-\alpha)ue^{-(\gamma
+\alpha)t}-\alpha uwe^{-(3\alpha+\gamma)t},\;\;\;\dot{w}=uve^{(\gamma
+\alpha)t}. \label{Chensys4}%
\end{equation}
Under the change of coordinates presented in Eq.(\ref{Chensys3}), the time
dependent first integral $H_{1}$ in Eq.(\ref{ChenInt1}) becomes autonomous%
\begin{equation}
H_{1}=u^{2}-2\alpha w. \label{ChenInt1i}%
\end{equation}
For the system (\ref{Chensys4}), the Jacobi's last multiplier can be taken as
the unity, hence the corresponding Poisson vector $\mathbf{J}_{1}$ is%
\begin{equation}
\mathbf{J}_{1}=\nabla H_{1}=\left(  2u,0,-2\alpha\right)  \label{ChenJ1}%
\end{equation}
while the Hamiltonian is
\begin{equation}
H_{2}=\frac{1}{2\alpha}\left[  \frac{1}{8\alpha}e^{-(3\alpha+\gamma)t}%
u^{4}+\frac{1}{2}(\gamma-\alpha)e^{-(\gamma+\alpha)t}u^{2}-\frac{1}{2}%
u^{2}we^{-(3\alpha+\gamma)t}\right]  -\frac{1}{4}v^{2}e^{(\gamma+\alpha)t}.
\label{ChenInt2i}%
\end{equation}
Using $H_{2}$, we compute the second Poisson matrix
\begin{equation}
\mathbf{J}_{2}=-\nabla H_{2}=\left(  \frac{e^{-(\gamma+\alpha)t}}{2\alpha
}\left(  e^{-2\alpha t}u\left(  w-\frac{u^{2}}{2\alpha}\right)  -\left(
\gamma-\alpha\right)  u\right)  ,\frac{e^{(\gamma+\alpha)t}}{2}v,\frac
{e^{-(3\alpha+\gamma)t}}{4\alpha}u^{2}\right)  . \label{ChenJ2}%
\end{equation}
Note that, this Poisson vector is similar to the one in Eq.(\ref{TJ2}) but it
has a finite value for large $t$ and $\gamma=-\alpha$. By substituting the
Poisson vectors $\mathbf{J}_{1}$ and $\mathbf{J}_{2}$ in Eqs.(\ref{ChenJ1})
and (\ref{ChenJ2}), and Hamiltonian functions $H_{1}$ and $H_{2}$ in
Eqs.(\ref{ChenInt1i}) and (\ref{ChenInt2i}), into Eq.(\ref{bi-Ham}), we arrive
bi-Hamiltonian structure of Chen system in Eq.(\ref{Chensys4}).

In \cite{HY}, a system
\begin{equation}
\dot{x}=\alpha(y-x),\text{ \ \ }\dot{y}=(\alpha-\gamma)x+\gamma y+\lambda
xz,\text{ \ \ }\dot{z}=xy-\beta z \label{ChenNew}%
\end{equation}
has been introduced by rescaling the cross term $xz$ in the second equation of
the Chen system (\ref{Chensys1}) by a real coefficient $\lambda$. We shall
show that the chaotic system (\ref{ChenNew}) is also bi-Hamiltonian. For
$\alpha=\beta=-\gamma$, it admits a time dependent integral
\begin{equation}
F_{1}=e^{2\alpha t}(x^{2}-y^{2}/2+\lambda z^{2}/2). \label{F1}%
\end{equation}
Upon making use of the transformation $(x,y,z)\rightarrow(u,v,w)$ given by
\begin{equation}
u=xe^{\alpha t},\text{ \ \ }v=ye^{\alpha t},\text{ \ \ }zw=ze^{\alpha t},
\label{Chen2NC}%
\end{equation}
the system (\ref{Chensys1}) turns out to be
\begin{equation}
\dot{u}=\alpha v,\text{ \ \ }\dot{v}=2\alpha u+\lambda uwe^{-\alpha t},\text{
\ \ }\dot{w}=uve^{-\alpha t}. \label{Chen2}%
\end{equation}
The system (\ref{Chen2}) is bi-Hamiltonian with Hamiltonian functions
\begin{equation}
F_{1}=u^{2}-v^{2}/2+\lambda w^{2}/2,\text{ \ \ }F_{2}=\alpha w-u^{2}e^{-\alpha
t}/2
\end{equation}
and the corresponding Poisson vectors%
\begin{equation}
\mathbf{J}_{1}=\nabla F_{1}=\left(  2u,-v,\lambda w\right)  ,\text{
\ \ }\mathbf{J}_{2}=-\nabla F_{2}=\left(  ue^{-\alpha t},0,-\alpha\right)  ,
\end{equation}
respectively.

\subsection{The Qi system}

Chaotic Qi system
\begin{align}
\dot{x} &  =\alpha(y-x)+yz,\label{LXsys1}\\
\dot{y} &  =\gamma x-xz-y,\nonumber\\
\dot{z} &  =xy-\beta z\nonumber
\end{align}
involves the addition of a cross product nonlinear term to the first equation
of the Lorenz system while retaining the linear feedback term in the second
equation of the Chen system (\ref{Chensys1}), see \cite{Qi}. When the
parameters of the Qi system (\ref{LXsys1}) satisfies $\alpha=\beta=1$, it
admits a time-dependent first integral%
\begin{equation}
H_{1}=e^{2t}\left(  \gamma x^{2}-y^{2}-(\gamma+1)z^{2}\right)  .\label{LXsys}%
\end{equation}
After the change of variables $\left(  x,y,z\right)  $ to $\left(
u,v,w\right)  $ given by
\begin{equation}
u=xe^{t},\;\;\;v=ye^{t},\;\;\;w=ze^{t},\label{LXsys2}%
\end{equation}
Qi system (\ref{LXsys1}) turns out to be non-autonomous
\begin{equation}
\dot{u}=v+vwe^{-t},\;\;\;\dot{v}=\gamma u-uwe^{-t},\;\;\;\dot{w}%
=uve^{-t}\label{LXsys3}%
\end{equation}
whereas the integral in Eq.(\ref{LXsys}) becomes time independent
\begin{equation}
H_{1}(u,v,w)=\gamma u^{2}-v^{2}-(\gamma+1)w^{2}.\label{LXsys4}%
\end{equation}
The Jacobi last multiplier $M=1$ and the Poisson vector
\begin{equation}
\mathbf{J}_{1}=\nabla H_{1}=\left(  2\gamma u,-2v,-2(\gamma+1)w\right)
\label{QiJ1}%
\end{equation}
with the Hamiltonian function
\begin{equation}
H_{2}=\frac{1}{2}w-\frac{1}{4(\gamma+1)}(u^{2}+v^{2})e^{-t}.\label{LXsys6}%
\end{equation}
The second Poisson vector is
\begin{equation}
\mathbf{J}_{2}=-\nabla H_{2}=\left(  \frac{u}{2(\gamma+1)}e^{-t},\frac
{v}{2(\gamma+1)}e^{-t},-\frac{1}{2}\right)  .\label{QiJ2}%
\end{equation}
By this we have achieved to write the bi-Hamiltonian structure of the system
(\ref{LXsys3}) where the Hamiltonian functions $H_{1}$ and $H_{2}$ are in
Eqs.(\ref{LXsys}) and (\ref{LXsys6}), and Poisson vectors in $\mathbf{J}_{1}$
and $\mathbf{J}_{2}$ are in Eqs.(\ref{QiJ1}) and (\ref{QiJ2}), respectively.

\section{Discussions and outlook}

In this study, we have presented bi-Hamiltonian structures of three
dimensional chaotic autonomous L\"{u}, modified L\"{u}, $T$, Chen and Qi
systems for some certain parameters that they involve. While achieving this,
two characteristics of three dimensional Hamiltonian systems were essential.
The first one is that, in three dimensions, a Hamiltonian system is mutually
bi-Hamiltonian \cite{GuNu93}, and the second one is that, in three dimensions,
a dynamical system is bi-Hamiltonian if and only if PDE in Eq.(\ref{JLM}) has
a non-trivial solution \cite{Gao}.

\section*{Acknowledgement}

We are extremely grateful to Professor H. G\"{u}mral for his valuable inputs
and enlightening discussions. PG gratefully acknowledge support from Professor
G. Rangarajan and IISC Mathematics Department where some part of the work was
performed. OE is grateful to Professor Z. Soyu\c{c}ok for his valuable suggestions.


\begin{thebibliography}{99}                                                                                               %


\bibitem {AbMa78}R. Abraham and J.E. Marsden \emph{Foundations of mechanics}.
Reading, Massachusetts: Benjamin/Cummings Publishing Company, (1978).

\bibitem {AGZ}A. Ay, M G\"{u}rses and K Zheltukhin, \emph{Hamiltonian
equations in} $\mathbb{R}^{3}$, J.Math. Phys. \textbf{44} (12) (2003) 5688-5705.

\bibitem {CaMaPe93}P. Casati, F. Magri and M. Pedroni, \emph{Bihamiltonian
manifolds and Sato's equations,} Integrable Systems, The Verdier Memorial
Conference, Editors O.Babelon, P.Cartier, Y.Kosmann-Schwarzbach. Progress in
Mathematics \textbf{115}, Birkh\"{a}user, (1993), 251-272.

\bibitem {CU1}G. Chen and T. Ueta, \emph{Yet another chaotic attractor}, Int.
J. Bifurcat. Chaos \textbf{9} (1999) 1465-1466.

\bibitem {CV}S. C$\breve{e}$likovsk$\acute{y}$ and A. Van$\breve{e}\breve{c}%
$ek, \emph{Bilinear systems and chaos}, Kybernetika, \textbf{30} (1994),
403-424. ibid, \emph{Control Systems: From linear analysis to synthesis of
chaos}, Prentice Hall Int., London-New York-Tokyo 1996.

\bibitem {EsGhGuGu15}O. Esen, A. G. Choudhury, P. Guha, and H. G\"{u}mral,
\emph{Superintegrable Cases of Four Dimensional Dynamical Systems}. arXiv
preprint (2015) arXiv:1510.05480.

\bibitem {Radio}Z. Elhadj, \emph{Analysis of a new 3 dimensional quadratic
chaotic system}, Radioengineering \textbf{17} (1) (2008) 9-13

\bibitem {Gao}P. Gao, \emph{Hamiltonian structure and first integrals for the
Lotka-Volterra systems}, Phys. Letts. A, \textbf{273} (2000) No. 1-2, 86-96, .

\bibitem {GN}C. Gonera and Y. Nutku, \emph{Super-integrable Calogero-type
systems admit maximal number of Poisson structures}, Phys. Lett. A
\textbf{285} (2001) No. 5-6, 301-306.

\bibitem {CGK}A. Ghose Choudhury, P. Guha and B. Khanra, \emph{On the Jacobi
last multiplier, integrating factors and the Lagrangian formulation of
differential equations of the Painlev\'{e}-Gambier classification}, J. Math.
Anal. Appl. \textbf{360} (2009) No. 2, 651--664.

\bibitem {GC}P. Guha and A. Ghose Choudhury, \emph{On Planar and Non-planar
Isochronous Systems and Poisson Structures}, Int. J. Geom. Methods Mod. Phys.
\textbf{7} (2010) No. 7, 1115-1131.

\bibitem {Guha06}P. Guha, \emph{Quadratic Poisson structures and Nambu
mechanics}, Nonlinear Analysis \textbf{65}, (2006), 2025-2034.

\bibitem {GuNu93}H. G\"{u}mral and Y. Nutku, \emph{Poisson structure of
dynamical systems with three degrees of freedom}, J. Math. Phys, \textbf{34}
(12) (1993) 5691-5723.

\bibitem {Gum1}H. G\"{u}mral, \emph{Existence of Hamiltonian structure in 3D},
Adv. Dyn. Syst. Appl. \textbf{5} (2010), No. 2, 159-171.

\bibitem {Gor01}A. Goriely, \emph{Integrability and nonintegrability of
dynamical systems}. Advanced Series in Nonlinear Dynamics, \textbf{19}. World
Scientific Publishing Co., Inc., River Edge, NJ, 2001. xviii+415 pp.

\bibitem {HB1}B. Hernandez-Bermejo, \emph{New solutions of the Jacobi
equations for three-dimensional Poisson structures}. J. Math. Phys.
\textbf{42} (2001) No. 10, 4984--4996.

\bibitem {HB2}B. Hernandez-Bermejo, \emph{One solution of the 3D Jacobi
identities allows determining an infinity of them}. Phys. Lett. A \textbf{287}
(2001) No. 5-6, 371--378.

\bibitem {HB3}B. Hernandez-Bermejo, \emph{New solution family of the Jacobi
equations: Characterization, invariants, and global Darboux analysis}, Journal
of mathematical physics \textbf{48}(2), (2007), 022903.

\bibitem {HY}K. Huang and Q. Yang, \emph{Stability and Hopf bifurcation
analysis of a new system}, Chaos, Solitons $\&$ Fractals \textbf{39} (2009) 567-578.

\bibitem {Jac1}C.G.J. Jacobi, \emph{Sul principio dell'ultimo moltiplicatore,
e suo uso come nuovo principio generale di meccanica}, Giornale Arcadico di
Scienze, Lettere ed Arti \textbf{99} (1844) 129-146.

\bibitem {Jac2}C.G.J. Jacobi, \emph{Theoria novi multiplicatoris systemati
aequationum differentialium vulgarium applicandi}, J. Reine Angew. Math 27
(1844), 199-268, Ibid 29(1845), 213-279 and 333-376. Astrophys. Journal, 342
(1989) 635-638.

\bibitem {MaMo84}F. Magri and C. Morosi, \emph{A geometrical characterization
of integrable Hamiltonian systems through the theory of Poisson-Nijenhuis
manifolds}, Quaderno 19-1984, Univ. of Milan.

\bibitem {LaPi12}C. Laurent-Gengoux, A. Pichereau and P. Vanhaecke,
\emph{Poisson structures}, Springer Science \& Business Media. Vol.
\textbf{347} (2012).

\bibitem {LM}P. Lieberman and C.-M Marle, \emph{Symplectic Geometry and
Analytical Mechanics}, D. Reidel Publishing Company (1987).

\bibitem {Lorenz}E.N. Lorenz, \emph{Deterministic nonperiodic flows}, J. Atmos
Sci. \textbf{20} (1963) 130-134.

\bibitem {LuChen}J. L\"{u} and G. Chen, \emph{A new chaotic attractor coined},
Int. J. Bifurcation and Chaos, \textbf{12 }(2002) No.3, 659-661.

\bibitem {LuChen1}J. L\"{u}, G. Chen, and Z. Zhang, \emph{A compound structure
of a new chaotic attractor}, Int.J.Bifurcation and Chaos \textbf{14} (2002) 669-672.

\bibitem {Nambu}Y. Nambu, \emph{Generalized Hamiltonian Mechanics}, Phys. Rev.
D \textbf{7} (1973) 2405-2412.

\bibitem {NL1}M.C. Nucci and P.G.L. Leach, \emph{Jacobi's last multiplier and
Lagrangians for multidimensional systems.}\textit{arXiv:0709.3231v1}.

\bibitem {NL2}M.C. Nucci and P.G.L. Leach, \emph{Jacobi's last multiplier and
symmetries for the Kepler problem plus a linear story}, J. Phys. A: Math. Gen.
\textbf{37} (2004) 7743-7753.

\bibitem {OLV}P. J. Olver, \emph{Applications of Lie groups to differential
equations}, Springer Science and Business Media \textbf{107} (2000).

\bibitem {Qi}G. Qi, G. Chen, S. Du, Z. Chen, Z. Yuan, \emph{Analysis of a new
chaotic system}, Physica A: Stat. Mech. Appl. \textbf{352} (2005) 295-308.

\bibitem {Ross}O.E. R\"{o}ssler, \emph{An equation for continuos chaos}, Phys.
Letts. A \textbf{57} (1976) 397-398.

\bibitem {Ta}L. Takhtajan, \emph{On Foundation of the Generalized Nambu
Mechanics}, Comm. Math. Phys. \textbf{160} (1994) 295.

\bibitem {TeVe04}A. Te\u{g}men and A. Ver\c{c}in, \emph{Superintegrable
systems, multi-Hamiltonian structures and Nambu mechanics in an arbitrary
dimension}, International Journal of Modern Physics A, \textbf{19}(3), (2004), 393-409.

\bibitem {TeWiHaMiPoRo}P. Tempesta, P. Winternitz, J. Harnad, W. Miller Jr, G.
Pogosyan and M. Rodriguez, \emph{Superintegrability in classical and quantum
systems}. American Mathematical Soc. (2004).

\bibitem {T1}G. Tigan and D. Opris, \emph{Analysis of a 3D chaotic system},
Chaos, Solitons $\&$ Fractals \textbf{36}(5) (2008) 1315-1319.

\bibitem {CU2}T. Ueta and G Chen, \emph{Bifurcation analysis of Chen's
equation}, Int. J. Bifurcat. Chaos \textbf{10} (2000) 1917-1931.

\bibitem {wei83}A. Weinstein, \emph{The local structure of Poisson manifolds}
, Journal of differential geometry, \textbf{18}(3), (1983) 523-557.

\bibitem {Whi}E.T. Whittaker, \emph{A Treatise on the Analytical Dynamics of
Particles and Rigid Bodies}, Dover, New York, 1944.
\end{thebibliography}
\end{document}